\documentclass[preprint,showpacs,showkeys,amsmath,amssymb]{revtex4}
\usepackage{graphicx}
\usepackage{dcolumn}
\usepackage{epsfig}
\usepackage{bm}
\newcommand{\be}{\begin{eqnarray}}
\newcommand{\en}{\end{eqnarray}}
\newcommand{\ben}{\begin{eqnarray*}}
\newcommand{\enn}{\end{eqnarray*}}

\newcommand{\f}{\frac}

\newcommand{\bi}{\begin{itemize}}
\newcommand{\ei}{\end{itemize}}

\newcommand{\la}{\langle}
\newcommand{\ra}{\rangle}

\begin{document}
%%%%%%%%%%%%%%%%%%%%%%%%%%%%%%%%%%%%%%%%%%%%%%%%%%%%%%%%%%%%%%%%%%%%%%%%%%%%%%%%%%%%%%%%%%%%%%%%%%%%%%%%%%%%%%%%%%%%%%%%%%%%
\title{On two-dimensionalization of three-dimensional turbulence in shell models}
%%%%%%%%%%%%%%%%%%%%%%%%%%%%%%%%%%%%%%%%%%%%%%%%%%%%%%%%%%%%%%%%%%%%%%%%%%%%%%%%%%%%%%%%%%%%%%%%%%%%%%%%%%%%%%%%%%%%%%%%%%%%
\author{Sagar Chakraborty}
\email{sagar@nbi.dk}
\affiliation{Niels Bohr International Academy, Blegdamsvej 17, 2100 Copenhagen $\phi$, Denmark}
\author{Mogens H. Jensen}
\email{mhjensen@nbi.dk}
\affiliation{Niels Bohr Institute, Blegdamsvej 17, DK-2100 Copenhagen, Denmark}
\author{Amartya Sarkar}
\email{amarta345@bose.res.in}
\affiliation{Department of Theoretical Sciences, S.N. Bose National Centre for Basic Sciences, Saltlake, Kolkata 700098, India}
%%%%%%%%%%%%%%%%%%%%%%%%%%%%%%%%%%%%%%%%%%%%%%%%%%%%%%%%%%%%%%%%%%%%%%%%%%%%%%%%%%%%%%%%%%%%%%%%%%%%%%%%%%%%%%%%%%%%%%%%%%%%
\date{\today}
%%%%%%%%%%%%%%%%%%%%%%%%%%%%%%%%%%%%%%%%%%%%%%%%%%%%%%%%%%%%%%%%%%%%%%%%%%%%%%%%%%%%%%%%%%%%%%%%%%%%%%%%%%%%%%%%%%%%%%%%%%%%
\begin{abstract}
Applying a modified version of the Gledzer-Ohkitani-Yamada (GOY) shell model, the signatures of so-called two-dimensionalization effect of three-dimensional incompressible, homogeneous, isotropic fully developed unforced turbulence have been studied and reproduced. 
Within the framework of shell models we have obtained the following results: (i) progressive steepening of the energy spectrum with increased strength of the rotation, and, (ii) depletion in the energy flux of the forward forward cascade, sometimes leading to an inverse cascade.
The presence of extended self-similarity and self-similar PDFs for longitudinal velocity differences are also presented for the rotating 3D turbulence case.
\end{abstract}
%%%%%%%%%%%%%%%%%%%%%%%%%%%%%%%%%%%%%%%%%%%%%%%%%%%%%%%%%%%%%%%%%%%%%%%%%%%%%%%%%%%%%%%%%%%%%%%%%%%%%%%%%%%%%%%%%%%%%%%%%%%%
\pacs{47.27.–i, 47.27.Jv, 47.32.Ef}
\keywords{Turbulence, GOY Shell Model, Two-dimensionalization, Rotation}
%%%%%%%%%%%%%%%%%%%%%%%%%%%%%%%%%%%%%%%%%%%%%%%%%%%%%%%%%%%%%%%%%%%%%%%%%%%%%%%%%%%%%%%%%%%%%%%%%%%%%%%%%%%%%%%%%%%%%%%%%%%%
\maketitle
%%%%%%%%%%%%%%%%%%%%%%%%%%%%%%%%%%%%%%%%%%%%%%%%%%%%%%%%%%%%%%%%%%%%%%%%%%%%%%%%%%%%%%%%%%%%%%%%%%%%%%%%%%%%%%%%%%%%%%%%%%%%
\section{Introduction}
Rotation of a fluid has been used as a control parameter that can progressively make a 3D turbulent flow look like a quasi-2D or a 2D turbulent flow. This phenomenon --- known as two-dimensionalization of three-dimensional (3D) turbulence --- may be termed two-dimensionalization effect.
The phrase `look like' in this connection basically means that certain properties of 3D turbulence, such as wavenumber dependence of energy spectrum, direction of energy cascade {\it etc.}, change to such behaviors that are more recognized with a 2D turbulence flow; in
other words the flow would seem to have become `two-dimensionalised'.
In view of the fact that the dynamics of oceans, atmospheres, liquid planetary cores, fluid envelopes of stars and other bodies of astrophysical and geophysical interest do require an understanding of inherent properties of turbulence in the rotating frame of reference, the problem of two-dimensionalization is naturally of profound scientific importance; turbulence in rotating bodies is even of some industrial and engineering interest.
\\
Two-dimensionalization appears to be a subtle non-linear effect, which is distinctly different from Taylor-Proudman effect, as shown in the works of Cambon\cite{Cambon1}, Waleffe\cite{Waleffe} and others.
Numerical simulations\cite{Smith} indicate the initiation of an inverse cascade of energy with rapid rotation, a fact well supported by the experiments\cite{Baroud1,Morize1}.
Although recent experiments by Baroud {\it et al.}\cite{Baroud1,Baroud2} and Morize {\it et al.}\cite{Morize1,Morize2} have shed some light on the two-dimensionalization effect, the scaling of two-point statistics and energy spectrum in rotating turbulence remains a controversial topic. An energy spectrum $E(k)\sim k^{-2}$ has been proposed\cite{Zhou,Canuto} for rapidly rotating 3D turbulent fluid and this does seem to be validated by some experiments\cite{Baroud1,Baroud2} and numerical simulations\cite{Yeung,Hattori,Reshetnyak,Muller}. But some experiments\cite{Morize1} do not agree with this proposed spectrum. They predict steeper than $k^{-2}$ spectrum and this again appears to be supported from by results\cite{Yang,Bellet} and analytical results found using wave turbulence theory\cite{Galtier,Cambon2}.
\\
It is well known that structure functions for 3D\cite{Kolmogorov}, quasi-2D\cite{Sagar1} and 2D\cite{Sagar2, Sagar22} turbulences contain quite a lot of information about the respective flows; for instance the exact results for third-order structure functions serve as benchmarks for any theory of turbulence.
Recently\cite{Sagar3,Sagar4}, there has been an attempt to calculate such non-trivial results for rotating turbulent flows.
Based on those results, it has been argued\cite{Sagar5} that the presence of helicity cascade in the rotating flow would cause depletion in the forward cascade of energy that sometimes may lead to inverse cascade and that the exponent ($-m$) in the energy spectrum $E(k)\sim k^{-m}$ should lie between $-2$ to $-7/3$ for rapid rotation.
\\
In this paper, we shall use GOY shell model\cite{Gledzer, Ohkitani}, modified appropriately, to investigate these signatures of two-dimensionalization effect, the behaviour of the structure function and the status of extended self-similarity (ESS)\cite{Benzi} in the rotating flows.
\\
One may indeed ask why one needs another shell model after Hattori {\it et. al.}\cite{Hattori} already have proposed modified version of Sabra shell model \cite{sabra} a few years ago.
To answer this question, let us collect the main results of that model: i) the exponent ($-m$) of the energy spectrum in the inertial range changes from $-5/3$ to $-2$, ii) no inverse cascade is detected with the increase in rotation rate, and iii) the PDF's of the longitudinal velocity difference doesn't match with the experiments.
Studies of the last few years' research-literature on turbulence would reveal that 
investigations of two-dimensionalization effect are growing rapidly.
Neat experiments have confirmed that the exponent ($-m$) overshoots the value $-2$ quite comfortably as for instance shown in the experiments by Morize {\it et. al.}\cite{Morize1}.
Moreover, the fact that, some experiments and numerical simulations do show inverse cascade with increase in the rotation rate, motivates us to construct shell models that can mimick this effect.
As mentioned above, Hattori {\it et. al.}'s model finds PDF which mismatches with experiments and also, the model requires a fluctuating part in the rotation rate to obtain various results, while in experiments and simulations such effects are not known.
This again should motivate the need for another model.
Moreover, the numerical simulations performed in the present paper are mainly for decaying turbulence while the model of Hattori {\it et. al.}'s dealt with forced turbulence.
Hence, in this paper we try to consider another possible shell model that can mimic the signatures of the two-dimensionalization effect in details.

\section{Shell Models: A Brief Overview}
Shell models for fluids are, in practice, simplified representative versions of the Navier-Stokes (NS) equations;
but they do retain enough of the flavour of the parent equations making themselves handy testing grounds for many statistical properties of fluid turbulence.
In fact, shell models have been used to study statistical properties of turbulence in the past\cite{Bohr,Biferale} with a fair degree of success.
The fact that 3D turbulence still lacks solid understanding, can be related to the lack of complete theoretical characterization and explanation of the energy-cascade mechanism --- the process that spreads and sustains turbulence over wide range of scales.
The popularity of shell models is due to their usefulness in modeling this very energy-cascade mechanism. 
The other advantage of using shell models is that, being deterministic dynamical models, they can be studied by dint of faster and accurate numerical simulations.
This stems from the fact that the number of degrees of freedom (DOF) needed to reach high Reynolds numbers($Re$) is just moderately high as the number of DOF grows logarithmically in $Re$ as opposed to NS case, where number of DOF goes as $Re^{9/4}$.
 \\
In brief, shell models are a set of coupled nonlinear differential equations each labeled by index $n=0,1,2,...$, called the shell index:
\be
\left(\frac{d}{dt}+\nu {k^2}_n\right)u_n=k_n(NL)_n[u,u]+f_n\label{0a}
\en
where the complex dynamical variables $u_n$ represent the temporal evolution of velocity fluctuations over a wavelength $k_n$.
The wavenumbers $k_n$ are given by $k_n=k_0\lambda^n$ with $\lambda$, called the intershell ratio, usually set to 2 and $k_0$ being a reference wavenumber.
The forcing term $f_n$ is taken as time-independent and is usually restricted to a single shell: $f_n=f\delta_{nn^*}$.
Velocity evolution is thus followed over a set of logarithmically equispaced shells.
The nonlinear term, $(NL)_n[u,u]$, is so chosen that in some sense total energy, helicity and phase space volume are conserved as is typical of the nonlinear term in the NS equation.
Another usual practice is demanding locality of interactions (only between nearest and next nearest shells) in shell(Fourier) space, although this is not absolutely necessary.
The advantage in doing so is one gets rid of the so called sweeping effect (the direct coupling between inertial and integral scales) making shell models ideal to study nontrivial temporal properties of the energy cascade mechanism because the time fluctuations are not obscured by the large-scale sweeping effect.
This makes shell models rather approximate, quasi-Lagrangian representations of NS equations.
\\
The choice of $(NL)_n$ is not at all unique and thus quite a few shell models have been proposed in the literature.
Most popular of the lot being the so-called Gledzer-Ohkitani-Yamada (GOY) model\cite{Gledzer,Ohkitani}; another example being the sabra model\cite{sabra}.
We however will concentrate solely on the GOY model and modify it appropriately to incorporate the effect of rotation.
%%%%%%%
\section{The Model}
The equation of motion describing 3D turbulence are the Navier-Stokes equations. For an incompressible fluid in a rotating frame it reads as:
\be
\frac{\partial\mathbf{u}}{\partial t}+\mathbf{u}\cdot\mathbf{\nabla u}+2\mathbf{\Omega}\times\mathbf{u} &=& -\frac{1}{\rho}\mathbf{\nabla}p+\nu\nabla^2\mathbf{u}+\mathbf{f}\\
\mathbf{\nabla}\cdot\mathbf{u} &=& 0
\en
where $\mathbf{\Omega}$ is the angular velocity of system rotation, $\nu$ the kinematic viscosity, $\rho$ the density and $\mathbf{f}$ the external force.
%
%In absence of $\mathbf{f}$ we consider decaying turbulence.
%
The term $2\mathbf{\Omega}\times\mathbf{u}$ in the equation is due to the Coriolis force and thus is absent in the non-rotating case.
As an aside, it may be mentioned that such type of linear terms can also be seen in complex cascade models for magnetohydrodynamic turbulence\cite{biskamp}.
\\
In this paper, we have adopted the following strategy for the numerical investigations \cite{SSR1,SSR2}.
A specific form of GOY shell model for non-rotating decaying 3D turbulence is:
\be
\left[\f{d}{dt}+\nu k_n^2\right]u_n=ik_n\left[u_{n+2}u_{n+1}-\f{1}{4}u_{n+1}u_{n-1}-\f{1}{8}u_{n-1}u_{n-2}\right]^*
\label{1}
\en
As discussed before, this may be thought as a time evolution equation for complex scalar shell velocities $u_n(k_n)$ that depends on $k_n$ --- the scalar wavevectors labeling a logarithmic discretised Fourier space ($k_n=k_02^n$).
We choose: $k_0=1/6,\nu=10^{-7} \textrm{ and } n=1 \textrm{ to }22$.
The initial condition imposed is: $u_n=k_n^{1/2}e^{i\theta_n}$ for $n=1,2$ and $u_n=k_n^{1/2}e^{-k_n^2}e^{i\theta_n}$ for $n=3 \textrm{ to }22$ where $\theta_n\in[0,2\pi]$ is a random phase angle.
The boundary conditions are: $u_n=0$ for $n<1$ and $n>22$.
It is easy to notice that GOY model has certain similarities with the NS equation.
The nonlinear terms have dimension $[\text{velocity}]^2/[\text{length}]$ as is the case with the nonlinear term($\mathbf{u}\cdot\mathbf{\nabla u}$) in the NS equation.
In the inviscid limit ($\nu\rightarrow 0$), equation (\ref{1}) owns two conserved quantities {\it viz.,} $\sum_n|u_n|^2$ and $\sum_n(-1)^nk_n|u_n|^2$ which are seen on the same footing as energy and helicity respectively.
Apart from this and most importantly, this model displays multiscaling, {\it i.e.} it has been shown\cite{Jensen} that at inertial scales, the structure functions (may be naively defined for shell models as $\langle\vert u_n\vert^p\rangle$) display power law dependence on $k_n$ with non-trivial exponents: $\langle\vert u_n\vert^p\rangle\propto k_n^{-\zeta_p}\label{1a}$, where the exponents $\zeta_p$ have a non-trivial nonlinear dependence on the order $p$.
%
%
%In the inviscid limit ($\nu\rightarrow 0$), equation (\ref{1}) owns two conserved quantities {\it viz.,} $\sum_n|u_n|^2$ (energy) and $\sum_n(-1)^nk_n|u_n|^2$ (helicity).
%
\\
If the fluid is rotating then one may modify equation (\ref{1}) by adding a term $R_n=-i\left[\omega+(-1)^nh\right]u_n$ in the R.H.S.
This specific form of the term $R_n$ was originally introduced by Reshetnyak and Steffen\cite{Reshetnyak}.
$\omega$ and $h$ are real numbers.
It may be noted that this term, as is customary of Coriolis force term in NS equations, wouldn't add up to the energy.
The $(-1)^nh$ part in $R_n$ has been introduced to have non-zero mean level of helicity that otherwise has a stochastic temporal behaviour and zero mean level.
Therefore, the appropriate shell model for rotating 3D turbulent fluid is:
\be
\left[\f{d}{dt}+\nu k_n^2\right]u_n=ik_n\left[u_{n+2}u_{n+1}-\f{1}{4}u_{n+1}u_{n-1}-\f{1}{8}u_{n-1}u_{n-2}\right]^*-i\left[\omega+(-1)^nh\right]u_n\label{2}
\en
We fix $h=0.1$ in our numerical experiments and test for the following values: $\omega=0.01,0.1,1.0$ and $10.0$.
Later, we shall come back to the effect of changing $h$ while keeping $\omega$ fixed.
We found ourselves poor in computational resources when trying to simulate for higher $\omega$, say $100$. the practical problem is that higher the $\omega$, longer one has to run the simulation to get an appreciable --- say, 10-shell-wide --- inertial range.
We shall henceforth refer $\omega$ as rotation strength.
\\
All the data points reported here are averaged over 2000 independent initial conditions and the error-bars reported herein are the corresponding standard deviations.
%
%%%%%%%%%%%%%%%%%%%%%%%%%%%%%%%%%%%%%%%%%%%%%%%%%%%%%%%%%%%%%%%%%%%%%%%%%%%%%%%%%%%%%%%%%%%%%%%%%%%%%%%%%%%%%%%%%%%%%%%%%%
\begin{figure}[]
\begin{center}
\epsfxsize=8cm \epsfysize=5cm
\rotatebox{0}{\epsfbox{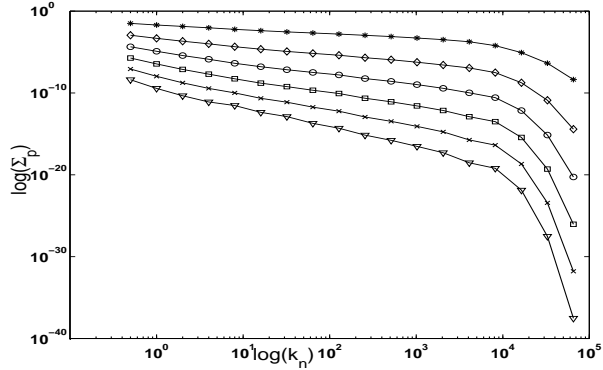}}
\end{center}
\caption[{Structure functions}]{A representative plot of structure functions $\Sigma_p$ vs. $k_n$, 
on log-log scale for $\omega=0.01, h=0.1$ for $\Sigma_p$ vs. $k_n$. From the topmost curve to the bottommost curve $p$ increases from $1$ to $6$. We plot for $n=3 \textrm{ to }20$.} 
\end{figure}
%%%%%%%%%%%%%%%%%%%%%%%%%%%%%%%%%%%%%%%%%%%%%%%%%%%%%%%%%%%%%%%%%%%%%%%%%%%%%%%%%%%%%%%%%%%%%%%%%%%%%%%%%%%%%%%%%%%%%%%%%%
%
Data have been recorded only after the energy cascade has stabilized and a nice the inertial range can be comfortably defined between shell $n=4 \textrm{ to }15$.
We have applied the slaved second order Adam-Bashforth scheme\cite{Pisarenko} to numerically
integrate equations (\ref{1}) and (\ref{2}).
\\
The $p$th order equal time structure function (see Fig. 1) for the model has been defined as:
\be
\Sigma_p(k_n)\equiv\left\la\left|\textrm{Im}\left[u_{n+1}u_{n}\left(u_{n+2}-\f{1}{4}u_{n-1}\right)\right]\right|^{\f{p}{3}}\right\ra\sim k_n^{-\zeta_p}
\en
to avoid period three oscillations\cite{Kadanoff}.
The energy spectrum has been defined as: \be E(k_n)=\Sigma_p(k_n)/k_n\sim k_n^{-m}.\en
The mean rate of dissipation of energy is, of course, \be\varepsilon=\left\la \sum_n\nu k_n^2 |u_n|^2\right\ra\en and flux through $n$th shell is calculated using the relation:
\be
&&\Pi_n\equiv\left\la-\f{d}{dt}\sum_{i=1}^{n}|u_i|^2\right\ra\\
\Rightarrow&&\Pi_n=\left\la-\textrm{Im}\left[k_nu_{n+1}u_{n}\left(u_{n+2}+\f{1}{4}u_{n-1}\right)\right]\right\ra
\en
For studying relative structure function scaling, the ESS scaling exponents are taken as \be \zeta_p^*\equiv\zeta_p/\zeta_3.\en
The exponents $m$, $\zeta_p$ and $\zeta_p^*$ have all been estimated for inertial ranges only.
\section{The Results}
In this section, we systematically present the results obtained by the numerical simulations
and based on these results we show, that indeed the two-dimensionalization effect is in 
fact mimicked by the present shell model.
\subsection{Signatures of two-dimensionalization}
%
%
%%%%%%%%%%%%%%%%%%%%%%%%%%%%%%%%%%%%%%%%%%%%%%%%%%%%%%%%%%%%%%%%%%%%%%%%%%%%%%%%%%%%%%%%%%%%%%%%%%%%%%%%%%%%%%%%%%%%%%%%%%
\begin{table*}[]
\begin{center}
\begin{tabular}{|c||c|c|c|c|c|}
\hline\hline
\multicolumn{6}{|l|}{\bf{Table 1: $\zeta_p$ for $p=1$ to $6$ for various rotation strengths.}}\\
\hline\hline
{$p$}&{$\zeta_p(\omega=0.00,h=0.0)$}&{$\zeta_p(\omega=0.01,h=0.1)$}&{$\zeta_p(\omega=0.10,h=0.1)$}&{$\zeta_p(\omega=1.00,h=0.1)$}&{$\zeta_p(\omega=10.0,h=0.1)$}\\\hline
1&0.37 $\pm$   0.0027&0.52  $\pm$  0.0086&0.63 $\pm$   0.0098&0.62 $\pm$   0.0067&0.66 $\pm$   0.0086\\
2&0.70  $\pm$  0.0062&0.95$\pm$    0.0182&1.1 $\pm$   0.0232&1.2 $\pm$   0.0161&1.2  $\pm$  0.0138\\
3&1.0  $\pm$  0.0127&1.3  $\pm$  0.0394&1.6 $\pm$   0.0455&1.7 $\pm$   0.0301&1.8 $\pm$   0.0197\\
4&1.3  $\pm$  0.0251&1.7 $\pm$   0.0712&2.0 $\pm$   0.0733&2.2 $\pm$   0.0490&2.4 $\pm$   0.0283\\
5&1.5 $\pm$   0.0454&2.0  $\pm$  0.1083&2.3 $\pm$   0.1017&2.7 $\pm$   0.0713&2.9 $\pm$   0.0402\\
6&1.8  $\pm$  0.0718&2.4 $\pm$   0.1470&2.7 $\pm$   0.1291&3.2 $\pm$   0.0953&3.4 $\pm$   0.0550\\
\hline\hline
\end{tabular}
\begin{tabular}{|c||c|c|c|c|c|}
\hline\hline
\multicolumn{6}{|l|}{\bf{Table 2: $\zeta_p^*\equiv\zeta_p/\zeta_3$ for $p=1$ to $6$ for various rotation strengths.}}\\
\hline\hline
{$p$}&{$\zeta_p^*(\omega=0.00,h=0.0)$}&{$\zeta_p^*(\omega=0.01,h=0.1)$}&{$\zeta_p^*(\omega=0.10,h=0.1)$}&{$\zeta_p^*(\omega=1.00,h=0.1)$}&{$\zeta_p^*(\omega=10.0,h=0.1)$}\\\hline
1&$0.37\pm0.0153$  &  $0.40\pm0.0480$  & $0.39\pm0.0553$& $ 0.36\pm0.0368$  & $0.37\pm0.0283$   \\
2& $0.70\pm0.0188 $ &$0.73\pm0.0576$    & $0.69\pm 0.0687$& $0.70\pm0.0463$  &  $0.67\pm 0.0335$  \\
3&$1.0\pm0.0253$  &  $1.0\pm0.0789$  & $1.0\pm0.0910$& $1.0\pm 0.0603$  & $1.0\pm0.0393 $  \\
4&  $1.3 \pm 0.0377$  & $1.3\pm 0.1106$ & $1.2\pm0.1188$& $1.3\pm0.0791$  & $1.3\pm0.0479$  \\
5& $1.5\pm0.0580$   &  $1.5\pm 0.1477$& $1.4\pm0.1472$& $1.6\pm0.1014$  & $1.6\pm0.0598$ \\
6&  $1.8\pm0.0844$  & $1.8\pm0.1865$ & $1.7\pm0.1746$& $1.9\pm0.1255$  & $1.9\pm0.0746$ \\
\hline\hline
\end{tabular}
\end{center}
\end{table*}
%%%%%%%%%%%%%%%%%%%%%%%%%%%%%%%%%%%%%%%%%%%%%%%%%%%%%%%%%%%%%%%%%%%%%%%%%%%%%%%%%%%%%%%%%%%%%%%%%%%%%%%%%%%%%%%%%%%%%%%%%%
%
%
One can clearly see, Fig.2 and Fig.3, that as the rotation strength increases, the energy spectrum becomes steeper and the slope monotonically increases from a value $\sim -5/3$ to a value of $\sim -7/3$; hence validating one of the two-dimensionalization effect's signatures.\\
Investigating the direction of the flux in the inertial range regime, we find (see Fig.4) that with the increase in rotation strength, then first the forward cascade rate starts decreasing and 
furthermore instances appear when at certain shells the flux direction reverses.
Again, the number of shells with such behavior increases as the rotation strength is enhanced; clearly suggesting a depletion in the rate of forward cascade.
Thus, yet another signature of two dimensionalization has appeared the shell model studies.\\
At this point, it must be appreciated how important the inclusion of term $-i(-1)^nh$ in equation (\ref{2}) is in responsible for the effect of depletion in the rate of forward cascade.
By setting mean level of helicity above zero, it is this very term that --- in accordance with the arguments\cite{Sagar4} that it is the helicity that is causing this signature of two dimensionalization effect to show up --- has empowered the model with the capacity to mimic the effect: attempts to see this very effect when setting $h=0$ fails.
To illustrate this, we fixed $\omega$ at $0.1$ and increased $h$ monotonically. As Fig.5 shows, for $h=0$ one doesn't get instances of negative flux; however, as $h$ increase, several of such instances can be observed. We feel that for our simulations $h=0.1$ is a good choice to arrive at various features of two-dimensionalization effect using this modified GOY shell model.
%
%
%%%%%%%%%%%%%%%%%%%%%%%%%%%%%%%%%%%%%%%%%%%%%%%%%%%%%%%%%%%%%%%%%%%%%%%%%%%%%%%%%%%%%%%%%%%%%%%%%%%%%%%%%%%%%%%%%%%%%%%%%%
\begin{figure}[h!]
\begin{center}
\epsfxsize=8cm \epsfysize=5cm
\rotatebox{0}{\epsfbox{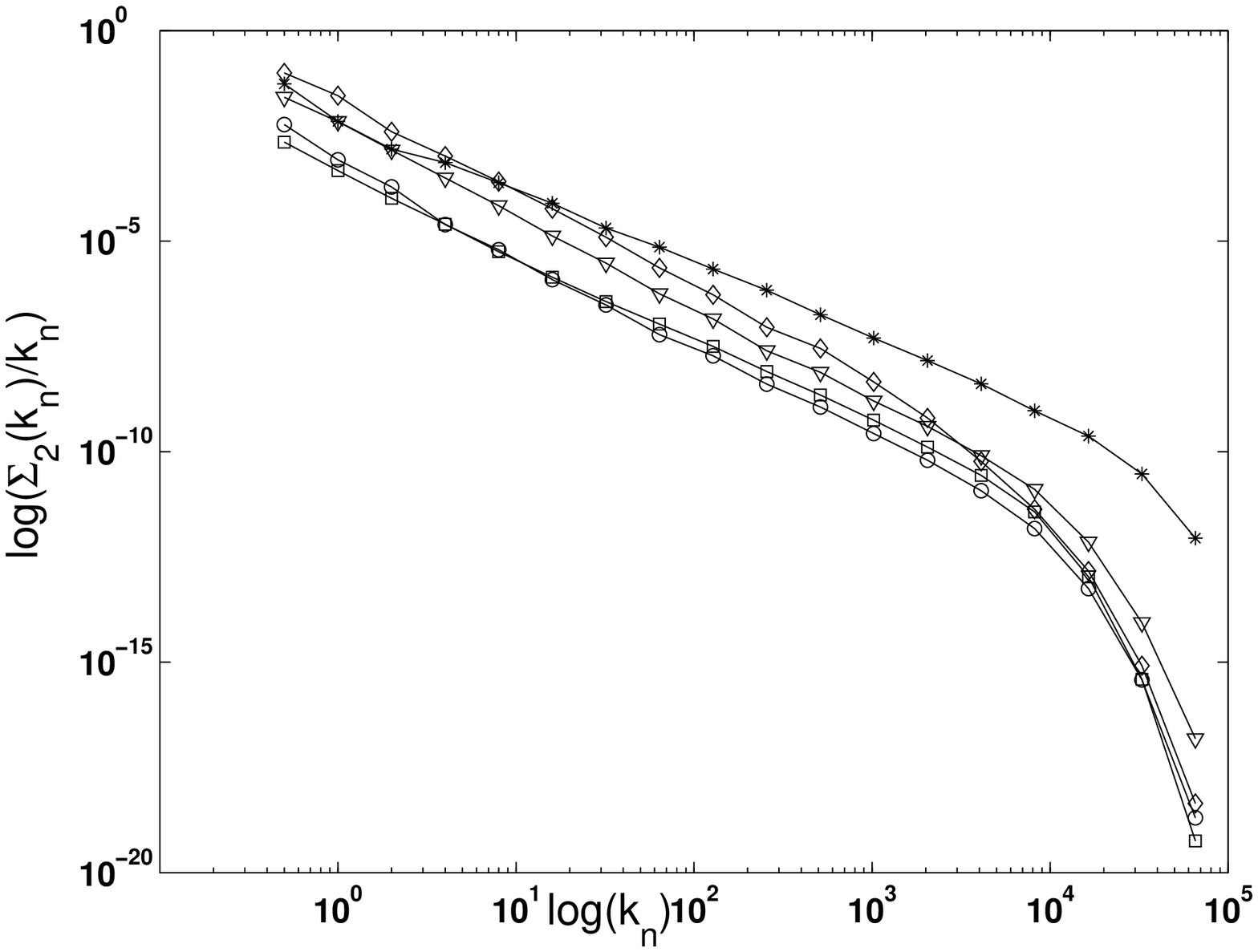}}
\rotatebox{0}{\epsfbox{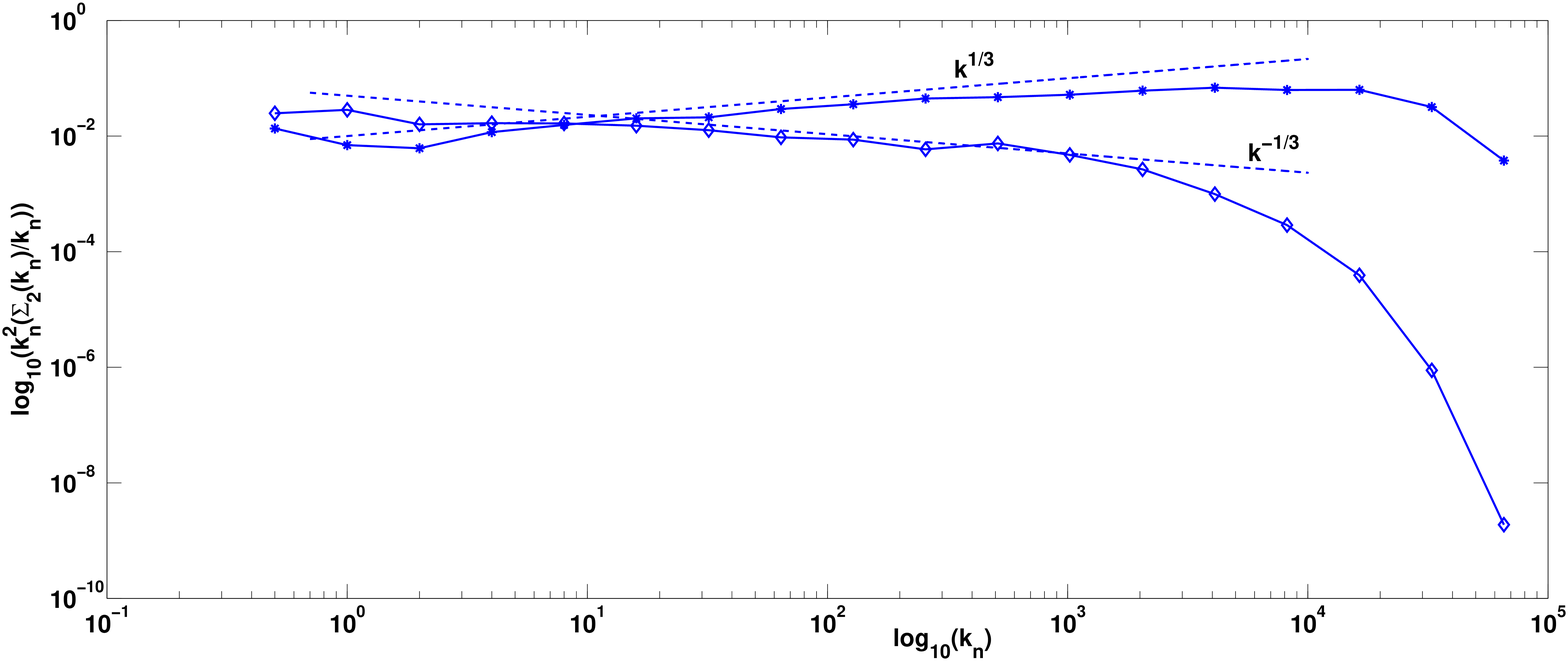}}
\end{center}
\caption[{}]{Energy spectra $E(k_n)$ vs. $k_n$ plotted in log-log plot. Asterisk is the marker for non-rotating case whereas square, triangle, circle and diamond respectively are the markers for $\omega=0.01$, $\omega=0.1$, $\omega=1.0$ and $\omega=10.0$ cases. We plot for $n=3 \textrm{ to }20$. For clarity, in an accompanying figure, we have also plotted compensated energy spectra $k_n^2E(k_n)$ vs. $k_n$ only for non-rotating and $\omega=10.0$ cases. One may note how the slope changes from $1/3$ to $-1/3$ with rotation as has been predicted\cite{Sagar3, Sagar4,Sagar5}.}\label{figspecslope}
\end{figure}
%%%%%%%%%%%%%%%%%%%%%%%%%%%%%%%%%%%%%%%%%%%%%%%%%%%%%%%%%%%%%%%%%%%%%%%%%%%%%%%%%%%%%%%%%%%%%%%%%%%%%%%%%%%%%%%%%%%%%%%%%%
%
%%%%%%%%%%%%%%%%%%%%%%%%%%%%%%%%%%%%%%%%%%
\begin{figure}[h!]
\begin{center}
\epsfxsize=8cm \epsfysize=5cm
\rotatebox{0}{\epsfbox{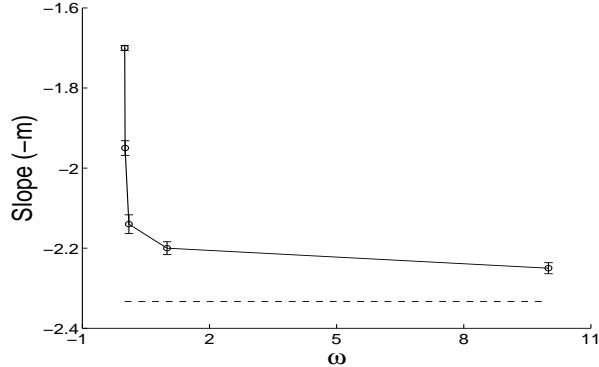}}
\end{center}
\caption[{}]{The slopes of the energy spectra (obtained from Fig.\ref{figspecslope}) plotted against the ``so-called'' rotation strength. The accompanying dashed line is the value $-7/3$ of the slope that has been predicted for very rapid rotation.}\label{figmvso}
\end{figure}
%%%%%%%%%%%%%%%%%%%%%%%%%%%%%%%%%%%%%%%%%%%%%%%%%%%%%%%%%%%%%%%%%%%%%%%%%%%%%%%%%%%%%%%%%%%%%%%%%%%%%%%%%%%%%%%%%%%%%%%%%%

%%%%%%%%%%%%%%%%%%%%%%%%%%%%%%%%%%%%%%%%%%%%%%%%%%%%%%%%%%%%%%%%%%%%%%%%%%%%%%%%%%%%%%%%%%%%%%%%%%%%%%%%%%%%%%%%%%%%%%%%%%
\begin{figure}
\begin{center}
\epsfxsize=8cm \epsfysize=4cm
%\rotatebox{0}{\epsfbox{flux1.eps}}
\rotatebox{0}{\epsfbox{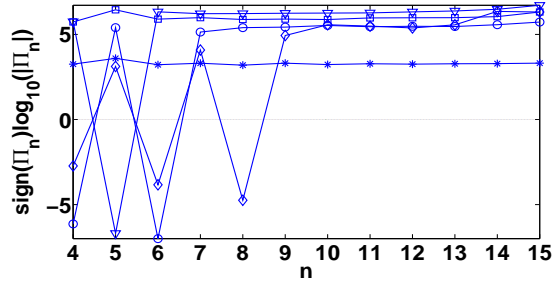}}
%\rotatebox{0}{\epsfbox{comfluxhel.eps}}
\end{center}
\caption[{Flux}]{Average flux of energy through $n$th shell vs. shell number $n$. The function $\Pi_n$  at the y-axis helps to plot all the curves distinctly in a single figure. Only the inertial range ($n=4$ to $15$) has been plotted. Markers are same as that for Fig. 2.}
\end{figure}
%%%%%%%%%%%%%%%%%%%%%%%%%%%%%%%%%%%%%%%%%%%%%%%%%%%%%%%%%%%%%%%%%%%%%%%%%%%%%%%%%%%%%%%%%%%%%%%%%%%%%%%%%%%%%%%%%%%%%%%%%%
%%%%%%%%%%%%%%%%%%%%%%%%%%%%%%%%%%%%%%%%%%%%%%%%%%%%%%%%%%%%%%%%%%%%%%%%%%%%%%%%%%%%%%%%%%%%%%%%%%%%%%%%%%%%%%%%%%%%%%%%%%
%\begin{figure}
%\begin{center}
%\epsfxsize=8cm \epsfysize=4cm
%\rotatebox{0}{\epsfbox{flux1.eps}}
%\rotatebox{0}{\epsfbox{comfluxhel.eps}}
%\end{center}
%\caption[{Flux}]{Average flux of energy through $n$th shell vs. shell number $n$. Only the inertial range ($n=4$ to $15$) has been plotted. Markers are same as that for fig-3.}
%\end{figure}
%%%%%%%%%%%%%%%%%%%%%%%%%%%%%%%%%%%%%%%%%%%%%%%%%%%%%%%%%%%%%%%%%%%%%%%%%%%%%%%%%%%%%%%%%%%%%%%%%%%%%%%%%%%%%%%%%%%%%%%%%%

%%%%%%%%%%%%%%%%%%%%%%%%%%%%%%%%%%%%%%%%%%%%%%%%%%%%%%%%%%%%%%%%%%%%%%%%%%%%%%%%%%%%%%%%%%%%%%%%%%%%%%%%%%%%%%%%%%%%%%%%%%
\begin{figure}
\begin{center}
\epsfxsize=8cm \epsfysize=4cm
%\rotatebox{0}{\epsfbox{flux1.eps}}
%\rotatebox{0}{\epsfbox{comflux.eps}}
\rotatebox{0}{\epsfbox{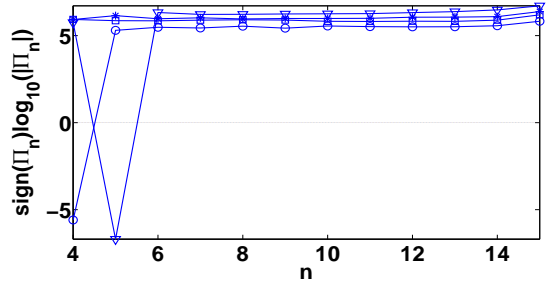}}
\end{center}
\caption[{Flux}]{Average flux of energy through $n$th shell vs. shell number $n$. Only the inertial range ($n=4$ to $15$) has been plotted. Asterisk, square, triangle and circle respectively are the markers for $h=0.0$, $h=0.01$, $h=0.1$ and $h=1.0$ cases. $\omega$ has been kept fixed at $0.1$.}
\end{figure}
%%%%%%%%%%%%%%%%%%%%%%%%%%%%%%%%%%%%%%%%%%%%%%%%%%%%%%%%%%%%%%%%%%%%%%%%%%%%%%%%%%%%%%%%%%%%%%%%%%%%%%%%%%%%%%%%%%%%%%%%%%
%
\subsection{Extended Self-Similarity}
The study of Extended Self-Similarity (ESS) in the shell model has also been revealing.
As can be seen in Fig.6 and Tables 1 and 2, the increase in the rotation strength is accompanied by a departure from the usual She-Leveque scaling.
But, the fact that at higher $p$, $\zeta_p$ seemingly becomes parallel to $p/2$ vs. $p$, is worth paying attention: This is in accordance with the direct numerical simulation (DNS) results\cite{Muller} and experimental results\cite{Baroud1}.
However, most interestingly is probably the observation that within the statistical error, $\zeta_p^*=\zeta_p/\zeta_3$ obtained for the rotating system via ESS coincides with that for the non-rotating ones.
Probably, this extends the ESS for 3D fluids even further by implying that rotation keeps ESS scaling intact, even though usual $\zeta_p$ changes owing to rotation.
Of course, only experiments and DNS can judge if this really is true for real fluid turbulence: The GOY shell, is after all just a model that remarkably well reproduces many characteristic features of turbulence by only using a fraction of computation power needed by DNS.
In this context, one might be well aware that some modified versions of GOY model invented to describe the distinguishing features of 2D turbulence have been shown to give spurious results \cite{Aurell}.
Thus, one always has to be careful while dealing with simplified models of turbulence.
%
%%%%%%%%%%%%%%%%%%%%%%%%%%%%%%%%%%%%%%%%%%%%%%%%%%%%%%%%%%%%%%%%%%%%%%%%%%%%%%%%%%%%%%%%%%%%%%%%%%%%%%%%%%%%%%%%%%%%%%%%%%
\begin{figure}[]
\begin{center}
\epsfxsize=8cm \epsfysize=8cm
\rotatebox{0}{\epsfbox{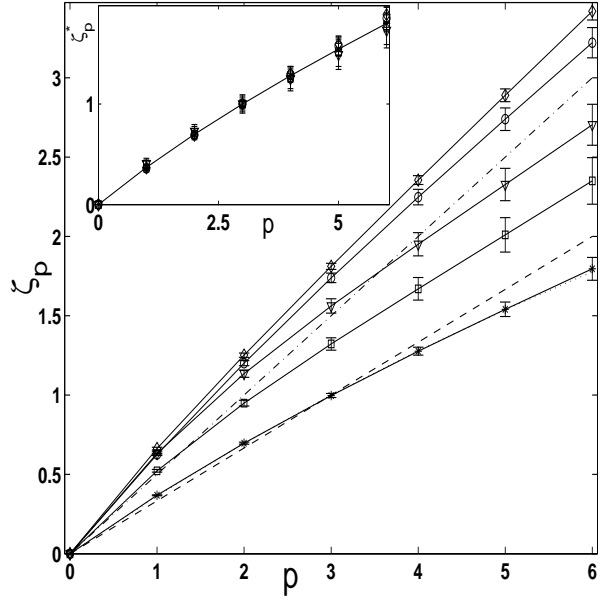}}
\end{center}
\caption[{ESS}]{$\zeta_p$ vs. $p$ plotted for the data in Table 1. Markers are same as that for Fig.-2. The dashed, the chain and the dotted lines are respectively for $\zeta_p=p/3$ (K41), $\zeta_p=p/2$ and $\zeta_p=p/9+2[1-(2/3)^{p/3}]$ (She-Leveque exponent\cite{She}). The dotted curve has almost been reproduced by non-rotating GOY model, as expected. This anomalous scaling is remarkably reproduced in the model dynamical system with limited number of degrees of freedom because its chaotic evolution exhibits temporal intermittency\cite{Jensen}. The inset is plot for $\zeta_p^*$ vs. $p$ plotted using the data of Table 2. All the connecting lines and the fractional values of $p$ are just aids for the eyes.}
\end{figure}
%%%%%%%%%%%%%%%%%%%%%%%%%%%%%%%%%%%%%%%%%%%%%%%%%%%%%%%%%%%%%%%%%%%%%%%%%%%%%%%%%%%%%%%%%%%%%%%%%%%%%%%%%%%%%%%%%%%%%%%%%
%

%
\subsection{Probability Distribution Functions}
We have also tried to see if we can get self-similar probability distribution function (PDF) for longitudinal velocity differences as has been reported in experiments\cite{Baroud1}. The GOY model is defined in $k$-space but we study the aforementioned PDF in real space obtained by using a sort of inverse Fourier transform\cite{Jensen2} of the form:
\begin{equation}
\label{field}
 {\vec v} ({\vec r},t)=\sum_{n=1}^{N}
 {\vec c}_n [u_n(t) e^{i {\vec k}_n\cdot {\vec r}}
+ c.\, c.].
\end{equation}
Here, the wavevectors are ${\vec k}_n ~=~ k_n {\vec e}_n$ where
${\vec e}_n$ is a unit vector in a random direction, for each shell
$n$ and ${\vec c}_n$ are unit vectors in random directions. We
ensure that the velocity field is incompressible, $\nabla\cdot{\vec
v} =0$, by constraining ${\vec c}_n \cdot {\vec e}_n =0,~\forall n
$. In our numerical computations we consider the vectors ${\vec
c}_n$ and ${\vec e}_n$ quenched in time but averaged over many
different realizations of these.
\\
Thus, when in the simulations energy cascade is stabilized, we use shell-velocities to find the real-space velocities following the aforementioned prescription.
From the velocity field obtained in this manner, one can easily construct longitudinal velocity difference as: \be \delta v_{||}\equiv[\vec{v}(\vec{r}+\vec{l})-\vec{v}(\vec{r})]\cdot\f{\vec{l}}{|\vec{l}|} .\en
We have chosen: $l=l_02^{1+3m}$ ($l_0\equiv 2\pi/k_N$) and have experimented for $m=0,1,\hdots,7$.
For a given separation $l$, we have calculated $\delta v_{||}$ for $10^5$ different $\mathbf{r}$ and normalized them by dividing $\delta{v}_{||}$ by their rms value.
%
%%%%%%%%%%%%%%%%%%%%%%%%%%%%%%%%%%%%%%%%%%%%%%%%%%%%%%%%%%%%%%%%%%%%%%%%%%%%%%%%%%%%%%%%%%%%%%%%%%%%%%%%%%%%%%%%%%%%%%%%%%
\begin{figure}[]
\begin{center}
\epsfxsize=8cm \epsfysize=6cm
\rotatebox{0}{\epsfbox{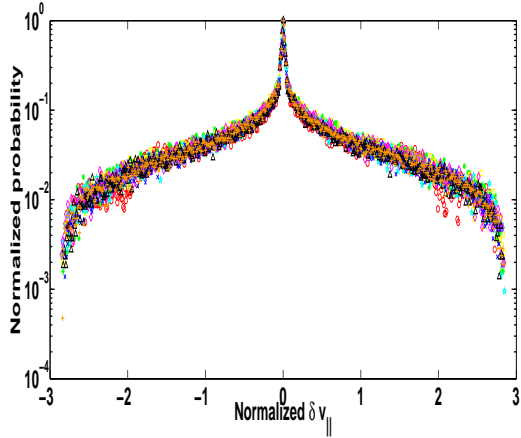}}
\end{center}
\caption[{ESS}]{({\it Color online}) Normalized probability vs. normalized longitudinal velocity difference for the unforced case. Longitudinal velocity difference, $\delta v_{||}\equiv[\vec{v}(\vec{r}+\vec{l})-\vec{v}(\vec{r})]\cdot\f{\vec{l}}{|\vec{l}|}$, is measured at varying distances. The x-axis basically is $\delta{v}_{||}$ divided by their rms value. $l=l_02^{1+3m}$ ($l_0\equiv 2\pi/k_N$). $m$ has been taken to be $0,1,\hdots,7$ and they respectively correspond to red circles, green asterisks, yellow squares, cyan stars, magenta diamonds, blue crosses, black triangles and orange pluses.}\label{figpdfdecay}
\end{figure}
%%%%%%%%%%%%%%%%%%%%%%%%%%%%%%%%%%%%%%%%%%%%%%%%%%%%%%%%%%%%%%%%%%%%%%%%%%%%%%%%%%%%%%%%%%%%%%%%%%%%%%%%%%%%%%%%%%%%%%%%%
%%%%%%%%%%%%%%%%%%%%%%%%%%%%%%%%%%%%%%%%%%%%%%%%%%%%%%%%%%%%%%%%%%%%%%%%%%%%%%%%%%%%%%%%%%%%%%%%%%%%%%%%%%%%%%%%%%%%%%%%%%
\begin{figure}[]
\begin{center}
\epsfxsize=8cm \epsfysize=6cm
\rotatebox{0}{\epsfbox{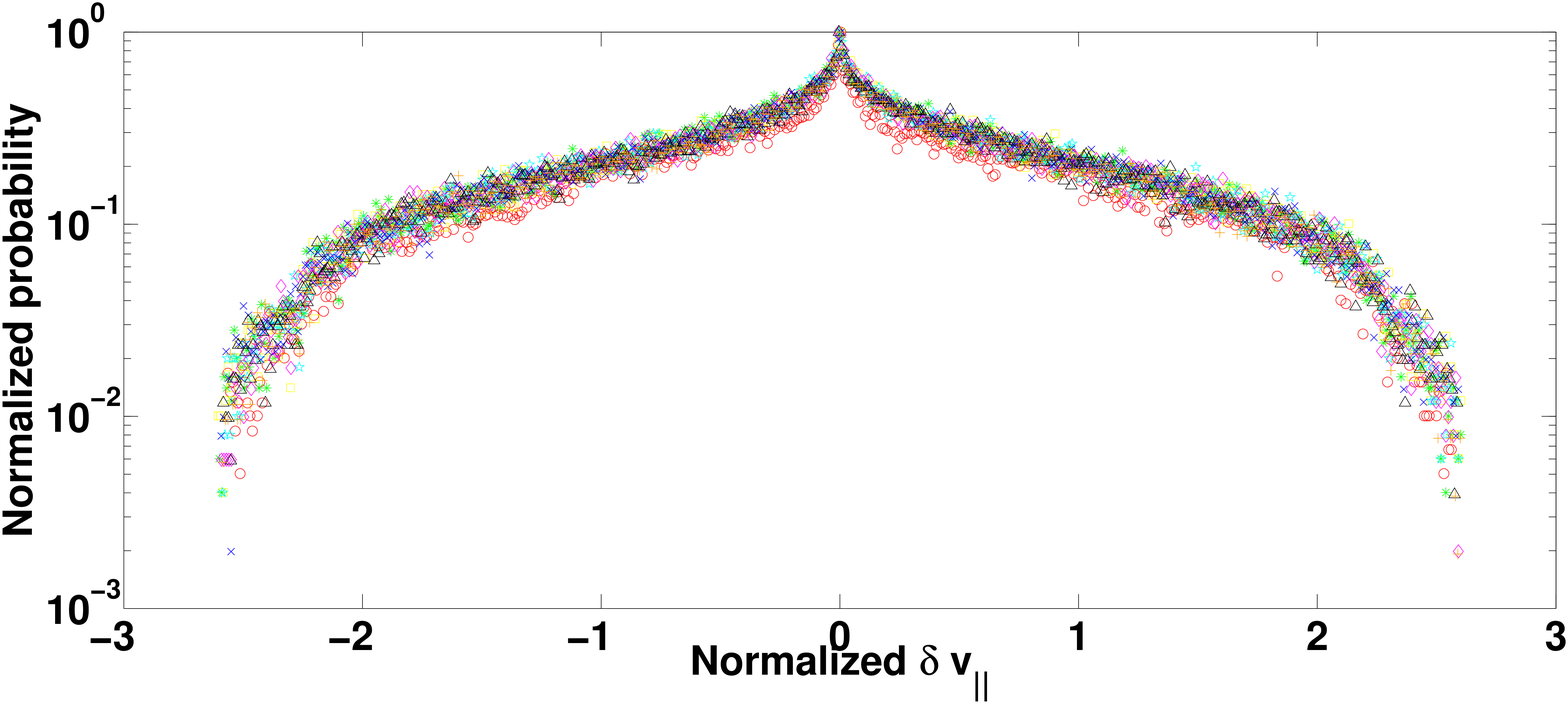}}
\end{center}
\caption[{ESS}]{({\it Color online}) Normalized probability vs. normalized longitudinal velocity difference for the forced case. Markers follow the convention used in figure (\ref{figpdfdecay}).}\label{figpdfforced}
\end{figure}
%%%%%%%%%%%%%%%%%%%%%%%%%%%%%%%%%%%%%%%%%%%%%%%%%%%%%%%%%%%%%%%%%%%%%%%%%%%%%%%%%%%%%%%%%%%%%%%%%%%%%%%%%%%%%%%%%%%%%%%%%
%
In Fig. \ref{figpdfdecay}, we present PDFs of normalized $\delta{v}_{||}$ for decaying rotating turbulence with an $\omega$-value equal to $10$. One may note that the PDFs are non-Gaussian but, when re-scaled appropriately, are fairly self-similar --- the plots for various separations (l) collapse on a single curve.
\\
However to make contact with experiments\cite{Baroud1}, we must also study a forced version of the model. Hence, we repeated our simulations for the case of rotating GOY shell model (given by equation (\ref{2})) with a forcing term in the R.H.S:
\be
\left[\f{d}{dt}+\nu k_n^2\right]u_n=ik_n\left[u_{n+2}u_{n+1}-\f{1}{4}u_{n+1}u_{n-1}-\f{1}{8}u_{n-1}u_{n-2}\right]^*-i\left[\omega+(-1)^nh\right]u_n+f\delta_{n,2}
\en
For this particular case, we chose $f=5\times10^{-3}(1+i)$ and we increased the total number of shells from $22$ to $24$. The corresponding PDFs for $\omega=10$, as given in Fig. (\ref{figpdfforced}), are non-Gaussian but also nicely self-similar. Thus, the rotating GOY model can reproduce even the self-similar feature of the PDFs quite impressively.
\section{Discussion and Conclusion}
Shell models have been successfully used to study statistical properties of turbulence by many authors (see Ref. (\cite{Bohr}) and Ref. (\cite{Biferale}) for details).
Most of the studies have dealt with the case of homogeneous and isotropic turbulence.
Hattori {\it et. al.} proposed a shell model for rotating turbulence.
Here we have attempted to improve their results by investigating two-dimensionalization effect
by using a modified version of GOY shell model.
Some results of the model are, no doubt, consistent with experiments and DNS.
\\
Concerning our main aim --- modeling the two-dimensionalization effect ---
one can always question the robustness of the obtained signatures because $i$) a scaling law for a single-component spectrum, though heavily used in literature, has poor meaning in the strongly anisotropic configuration which is relevant when passing from 3D-2D; different power laws can be found in terms of $k_z$, $k_{\bot}$ and $k$ in contrast to the 3D isotropic case, and $ii$) the inertial wave-turbulence theory is not consistent with an inverse cascade.
Actually in weak-wave turbulence, getting rid provisionally of helicity and polarization spectra, a two-component energy spectrum $e(k,\cos\theta)$ with $\cos\theta=k_z/\sqrt{(k^2_z+k^2_{\bot}})$ is found to be useful; if $E$ denotes the traditional spherically averaged spectrum, the anisotropic structure is one of the best ways to quantify all intermediate states from isotropic 3D (with $e=E(k)/(4\pi k^2)$) to 2D state (with $e=E(k_{\bot})/(2\pi k_{\bot})\delta(k_z)$).
Two-dimensional trends are therefore linked to a preferred concentration of spectral energy towards the transverse wave-plane $k_z=0$.
This concentration, however, does not necessarily yield an inverse cascade\cite{Galtier}.
A reasonable suggestion, in the light of this discussion, would be that in the shell model for rotating turbulence $k$ should be interpreted as $k_{\bot}$.
\\
In the closing, it may be concluded that this study has put the equation (\ref{2}) as a very good shell model for the rotating 3D turbulent flows; after all, it explains the observed signatures of the two-dimensionalization effect closely.
Probably, this model and the model due to Hattori {\it et. al.} can together model the rotating turbulence in a simple but effective manner.
\acknowledgements
SC thanks Prof. J.K. Bhattacharjee and his colleagues in S.N.B.N.C.B.S., Kolkata. He is grateful to S. Bhattacharjee, S. S. Ray and P. Perlekar for fruitful discussions. 
%

%%%%%%%%%%%%%%%%%%%%%%%%%%%%%%%%%%%%%%%%%%%%%%%%%%%%%%%%%%%%%%%%%%%%%%%%%%%%%%%%%%%%%%%%%%%%%%%%%%%%%%%%%%%%%%%%%%%%%%%%%
\end{document}